\documentstyle[amssymb,psfig,12pt]{article}
\begin{document}

\title{Quantum collision states for positive charges in an octahedral cage}
\author{R. Vilela Mendes\thanks{%
e-mail: vilela@cii.fc.ul.pt} \\
{\small Laborat\'{o}rio de Mecatr\'{o}nica,}\\
{\small DEEC, Instituto Superior T\'{e}cnico}\\
{\small Av. Rovisco Pais, P1096 Lisboa Codex, Portugal}}
\date{}
\maketitle

\begin{abstract}
One-electron energy levels are studied for a configuration of two positive
charges inside an octahedral cage, the vertices of the cage being occupied
by atoms with a partially filled shell. Although ground states correspond to
large separations, there are relatively low-lying states with large
collision probabilities. Electromagnetic radiation fields used to excite the
quantum collisional levels may provide a means to control nuclear reactions.
However, given the scale of the excitation energies involved, this mechanism
cannot provide an explanation for the unexplained ``cold fusion'' events.
\end{abstract}


\section{Introduction}

Inducing collision or near-collision states of like charged particles means
overcoming the strong Coulomb repulsion at small distances, a very
challenging task indeed. However, to master these events would have
plentiful of potential applications in chemical processes and in the control
of nuclear reactions. The traditional way to overcome the Coulomb repulsion
is by endowing one or both particles with sufficient kinetic energy, either
by acceleration or by thermal means. There are however subtler means to
achieve this goal, which must obviously involve some other particles of
opposite charge.

Classical configurations of particles of different charges in a close
neighborhood are unstable and cannot provide a steady shielding effect.
Turning to quantum mechanics, one also knows that static shielding effects
of ground state atomic orbitals, operative at atomic scales, do not provide
adequate shielding at nuclear scale distances. However, quantum mechanics
has some other subtler effects, namely the appearance of well-defined
excited levels with wave functions located around some of the unstable
classical orbits ({\it scars }\cite{Heller}). A particularly interesting
case corresponds to configurations related to unstable saddle point of the
potential ({\it saddle scars }\cite{Vilela1}). As opposed to the classical
case, where it is always difficult to make use of unstable orbits (except in
a few low dimensional cases\cite{Ott}), when these orbits have a scarred
quantum counterpart, these states may be easily addressed and maintained by
resonant excitation at the appropriate energy.

In this paper one computes the electronic states for a configuration of two
positive charges in a octahedral cage. As expected, a static calculation
leads to a lowest energy state with widely separated positive charges.
However when a dynamical degree of freedom is included for the positive
charges, one finds relatively low-lying excited states with a large
collision probability. These calculations are described in the next section.
Finally, the last section is dedicated to the experimental implications of
the results for the control of nuclear reactions and to a discussion of
other related and unrelated results concerning similar questions.

\section{Two charges in an octahedral cage}

One considers two positive charges inside an octahedral cage. In the six
vertices of the octahedral cage lie atoms with a filled closed shell and
some partially filled $d$-levels. The atoms at the vertices are assumed to
be fixed and the two positive charges are symmetrically placed on the
diagonal (see Fig.1).

\begin{figure}[htb]
\begin{center}
\psfig{figure=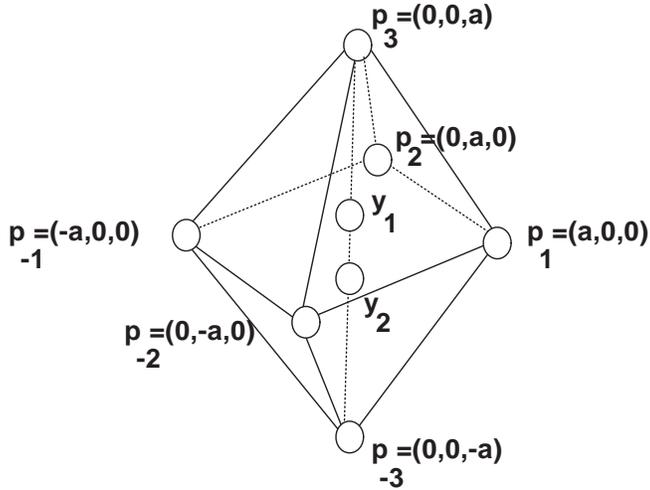,width=9truecm}
\end{center}
\caption{Two charges in an octahedral cage}
\end{figure}

Two distinct cases are studied. In the first the positive charges are
assumed to be fixed and in the second they are allowed to move symmetrically
along the diagonal.

The one-electron energy levels inside the octahedral cage are studied
assuming, as an approximation, that the available orbitals are:

(i) The $d$-orbitals centered at the vertices having $m=0$ projection
towards the interior of the octahedron, namely 
\begin{equation}
\chi _{\pm i}^{(d)}=N_{1}r_{2}^{-3/2}\left\{ 3\left( \frac{\left( 
\overrightarrow{x}-\overrightarrow{p_{\pm i}}\right) \cdot e_{i}}{\left| 
\overrightarrow{x}-\overrightarrow{p_{\pm i}}\right| }\right) -1\right\}
g_{nd}\left( \frac{\left| \overrightarrow{x}-\overrightarrow{p_{\pm i}}%
\right| }{r_{2}}\right)  \label{2.1}
\end{equation}
$\overrightarrow{x}$ being the coordinate of the electron, $\overrightarrow{%
p_{i}}$ the coordinates of the vertices and $g_{nd}$ a radial function for $%
d $-states. The normalization constant $N_{1}$ is fixed by normalizing the
function in the volume of the octahedron.. The effective radius $r_{2}$
takes into account the shielding effect of the closed shell.

(ii) The $s$-orbitals centered at the positive charges $y_{1}$ and $y_{2}$%
\begin{equation}
\chi _{j}^{(s)}=N_{0}\exp \left( -\frac{\left| \overrightarrow{x}-%
\overrightarrow{y_{j}}\right| }{r_{1}}\right)  \label{2.2}
\end{equation}
with the same prescription as above for the computation of the normalization
constant $N_{0}$.

When the positive charges are allowed to move symmetrically along the
diagonal one also uses a basis of Legendre polynomials in the $z$-coordinate 
\begin{equation}
z=\left| \overrightarrow{y_{1}}-\overrightarrow{y_{2}}\right|  \label{2.3}
\end{equation}
With $N$ polynomials one has a (non-orthogonal) basis of $8N$ functions.

This setting is recognized to be too simple to obtain accurate numerical
values of energy levels for a realistic situation of positive charges
confined in a metallic lattice. Nevertheless it seems to be qualitatively
correct and to provide a reasonable control on the nature of the excited
states, which is the main objective.

The Hamiltonian is 
\begin{eqnarray}
aH &=&-\frac{\hbar ^{2}}{2m_{e}a}\Delta _{x}-\sum_{j=1}^{2}\frac{\hbar ^{2}}{%
2M_{+}a}\Delta _{y_{j}}-\frac{e^{2}}{4\pi \varepsilon _{0}}\sum_{k=-3}^{3}%
\frac{Z_{eff}}{\left| \overrightarrow{x}-\overrightarrow{p_{k}}\right| }
\label{2.4} \\
&&+\frac{e^{2}}{4\pi \varepsilon _{0}}\left( -\sum_{j=1}^{2}\frac{1}{\left| 
\overrightarrow{x}-\overrightarrow{y_{j}}\right| }+\frac{1}{\left| 
\overrightarrow{y_{1}}-\overrightarrow{y_{2}}\right| }+\sum_{j=1,2;k=-3...3}
\frac{Z_{eff}}{\left| \overrightarrow{y_{j}}-%
\overrightarrow{p_{k}}\right| }\right)   \nonumber
\end{eqnarray}
all distances being measured in units of a reference value $a$ (one half the
octahedron diagonal). Eq.(\ref{2.4}) may be rewritten as 
\begin{eqnarray}
H^{^{\prime }} &=&\frac{4\pi \varepsilon _{0}a}{e^{2}}H=-\frac{0.264}{a(%
\stackrel{\circ }{A} )}\left( \Delta _{x}+\sum_{j=1}^{2}\frac{m_{e}}{M_{+}%
}\Delta _{y_{j}}\right) -\sum_{k=-3}^{3}\frac{Z_{eff}}{\left| 
\overrightarrow{x}-\overrightarrow{p_{k}}\right| }  \label{2.5} \\
&&-\sum_{j=1}^{2}\frac{1}{\left| \overrightarrow{x}-\overrightarrow{y_{j}}%
\right| }+\frac{1}{\left| \overrightarrow{y_{1}}-\overrightarrow{y_{2}}%
\right| }+\sum_{j=1,2;k=-3...3} \frac{Z_{eff}}{\left| 
\overrightarrow{y_{j}}-\overrightarrow{p_{k}}\right| }  \nonumber
\end{eqnarray}
where, $m_{e}$ being the electron mass, the relation 
\begin{equation}
\frac{\hbar ^{2}}{2m_{e}a}\frac{4\pi \varepsilon _{0}}{e^{2}}=\frac{0.264}{a(%
\stackrel{\circ }{A})}  \label{2.6}
\end{equation}
has been used, with $a$ expressed in Angstroms.

\subsection{Static charges}

Here one studies the energy spectrum as a function of the separation of the
two positive static charges located at $y_{1}=\left( 0,0,l\right) $ and $%
y_{2}=\left( 0,0,-l\right) $. The rescaled Hamiltonian $H^{^{\prime }}$ in
the static case may be split into two pieces $H^{^{\prime }}=H_{x}+H_{0}$%
\begin{equation}
\begin{array}{lll}
H_{x}\left( l\right)  & = & -\frac{0.264}{a(\stackrel{\circ }{A})}\Delta
_{x}-\sum_{k=-3}^{3}\frac{Z_{eff}}{\left| \overrightarrow{x}-\overrightarrow{%
p_{k}}\right| }-\sum_{j=1}^{2}\frac{1}{\left| \overrightarrow{x}-%
\overrightarrow{y_{j}}\right| } \\ 
H_{0}\left( l\right)  & = & \frac{1}{2l}+\sum_{j=1,2;k=-3...3}
 \frac{Z_{eff}}{\left| \overrightarrow{y_{j}}-\overrightarrow{p_{k}}%
\right| }
\end{array}
\label{2.7}
\end{equation}
The $l-$dependent energy spectrum of $H_{x}$ in the basis $\left\{ \chi
_{j}^{(s)},\chi _{\pm i}^{(d)}\right\} $ (Eqs.(\ref{2.1})-(\ref{2.2})) 
\[
\lambda _{1}\left( l\right) ,\lambda _{2}\left( l\right) ,\cdots ,\lambda
_{8}\left( l\right) 
\]
has been computed. In Fig.2 one shows the energy of 1, 2 and 16 electrons
(not corrected for electron-electron interactions), that is 
\begin{equation}
\begin{array}{lll}
E_{1}\left( l\right) = & \lambda _{1}\left( l\right) +H_{0}\left( l\right) 
&  \\ 
E_{K}\left( l\right) = & \sum_{i=1}^{K/2}2\lambda _{i}\left( l\right)
+H_{0}\left( l\right)  & \qquad (K=2,16)
\end{array}
\label{2.8}
\end{equation}
as a function of $l$. The last plot in Fig.2 shows $H_{0}\left( l\right) $.
All results drawn in the figures of this paper correspond to $Z_{eff}=10$
and $a=2.05\stackrel{\circ }{A} $. Qualitatively similar results are
obtained for other values of the constants.

\begin{figure}[htb]
\begin{center}
\psfig{figure=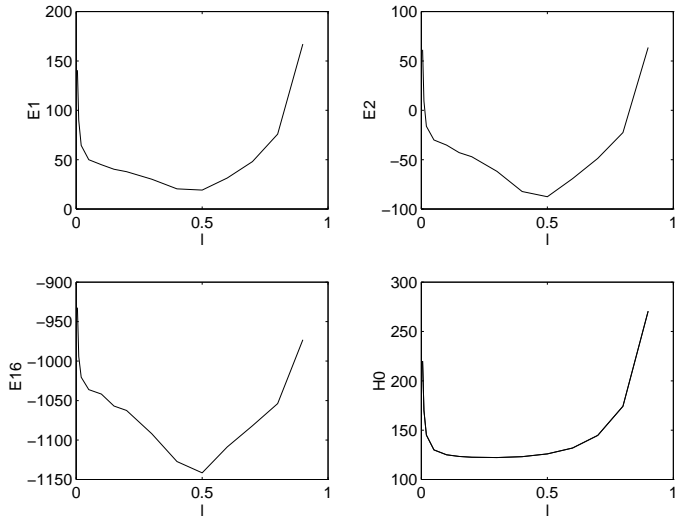,width=9truecm}
\end{center}
\caption{Energy of 1, 2 and 16 electron configurations and static energy}
\end{figure}

One sees that, for all the electron configurations, the minimum energy
occurs for a large separation of the positive charges at $y_{1}$ and $y_{2}$. 
For comparison the energy for two electrons and two positive charges in
empty space (an isolated molecule) was also computed for the same
parameters. The result is shown in Fig.3. One concludes that the octahedral
cage and the $d$ electron levels have the effect of increasing rather than
decreasing the separation of the positive charges. This conclusion is
similar to the result of Sun and Tom\'{a}nek\cite{Sun} that, using a
density-functional calculation, concluded that the equilibrium distance
between two deuterium atoms in a palladium lattice is larger than the gas
value.

\begin{figure}[htb]
\begin{center}
\psfig{figure=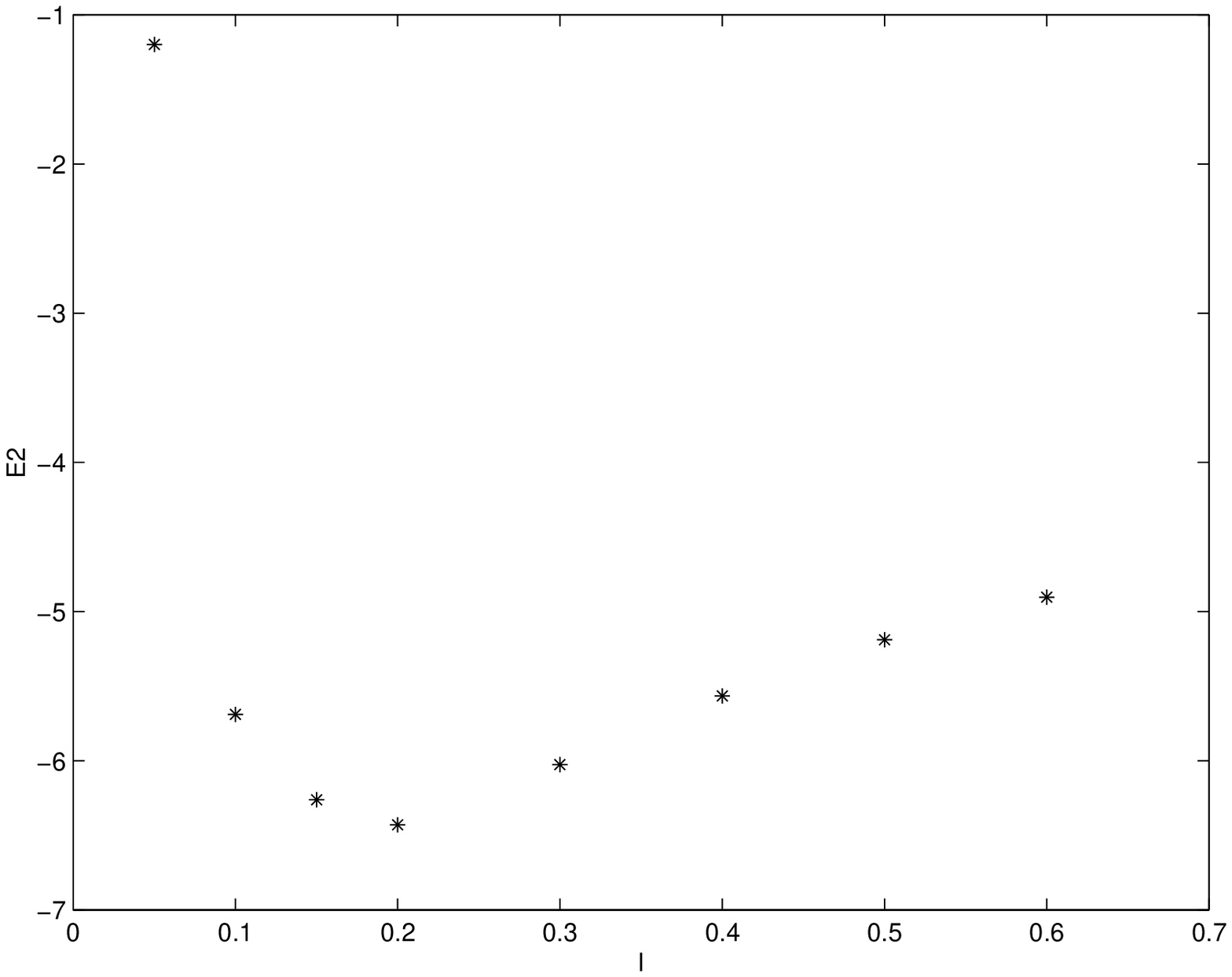,width=9truecm}
\end{center}
\caption{Energy of an isolated molecule}
\end{figure}

The fact that the minimum energy is obtained for very large separations,
does not preclude the existence of low lying excited states, with large
occupation probabilities at zero separation. To clarify this point one needs
to solve the problem allowing, at least, the two charges to move along the $%
z-$axis.

Why a dynamical calculation may provide information qualitatively different
from the static case is easy to understand. A static calculation is
equivalent to constrain the wave functions in the relative coordinate to be
a delta-function at $z=2l$. If $l$ is small this means that the two positive
charges are all the time very close to each other, independently of the
position of the electrons. For the dynamic case the situation is different,
because one may have high probabilities at $z=0$ if they correspond to
configurations where there is also a high probability for the electrons to
be near the origin of the octahedron.

Of course, situations of this type are energetically favorable only if the
overall potential of positive and negative charges has local minima near $%
z=0 $. In particular, if these are saddle points of the potential, it is
known that the unstable classical equilibrium configurations manifest
themselves as well defined quantum states \cite{Vilela1}.

\subsection{Dynamic charges}

Here one allows the positive charges at $y_{1}$ and $y_{2}$, to move
symmetrically along the $z-$axis. By allowing just one additional dynamical
degree of freedom, the problem is kept computationally simple, while at the
some time a large enough basis may be used for the $z$ degree of freedom. A
basis of $8$ Legendre polynomials $P_{n}(z)$ in the $z-$coordinate is used.
Together with the states defined in Eqs.(\ref{2.1})-(\ref{2.2}) the basis
has now 64 states. The one-electron spectrum that is obtained is plotted in
the upper plot of Fig.4. The lower plot shows the value of the projected
squared wave-function $\left| \psi \right| ^{2}$ at $z=0$. $M_{+}$ is taken
to be the deuterium mass and for $Z_{eff}$ and $a$ the values are the same
as in the static case.

\begin{figure}[htb]
\begin{center}
\psfig{figure=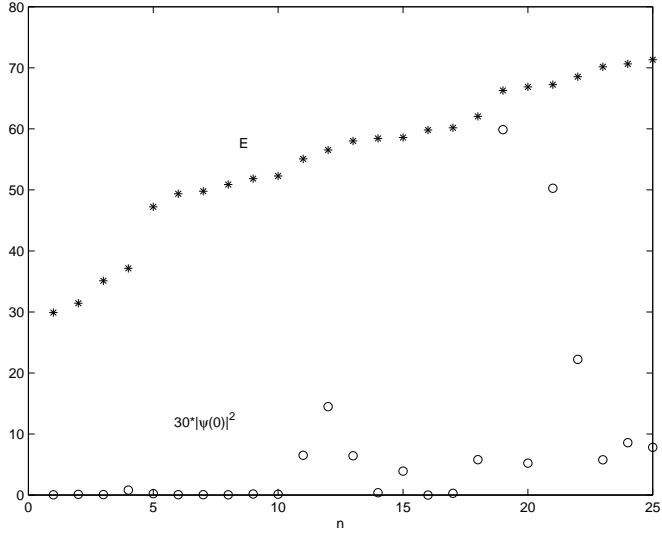,width=9truecm}
\end{center}
\caption{One-electron spectrum and projected squared wave-function at $z=0$}
\end{figure}

\begin{figure}[htb]
\begin{center}
\psfig{figure=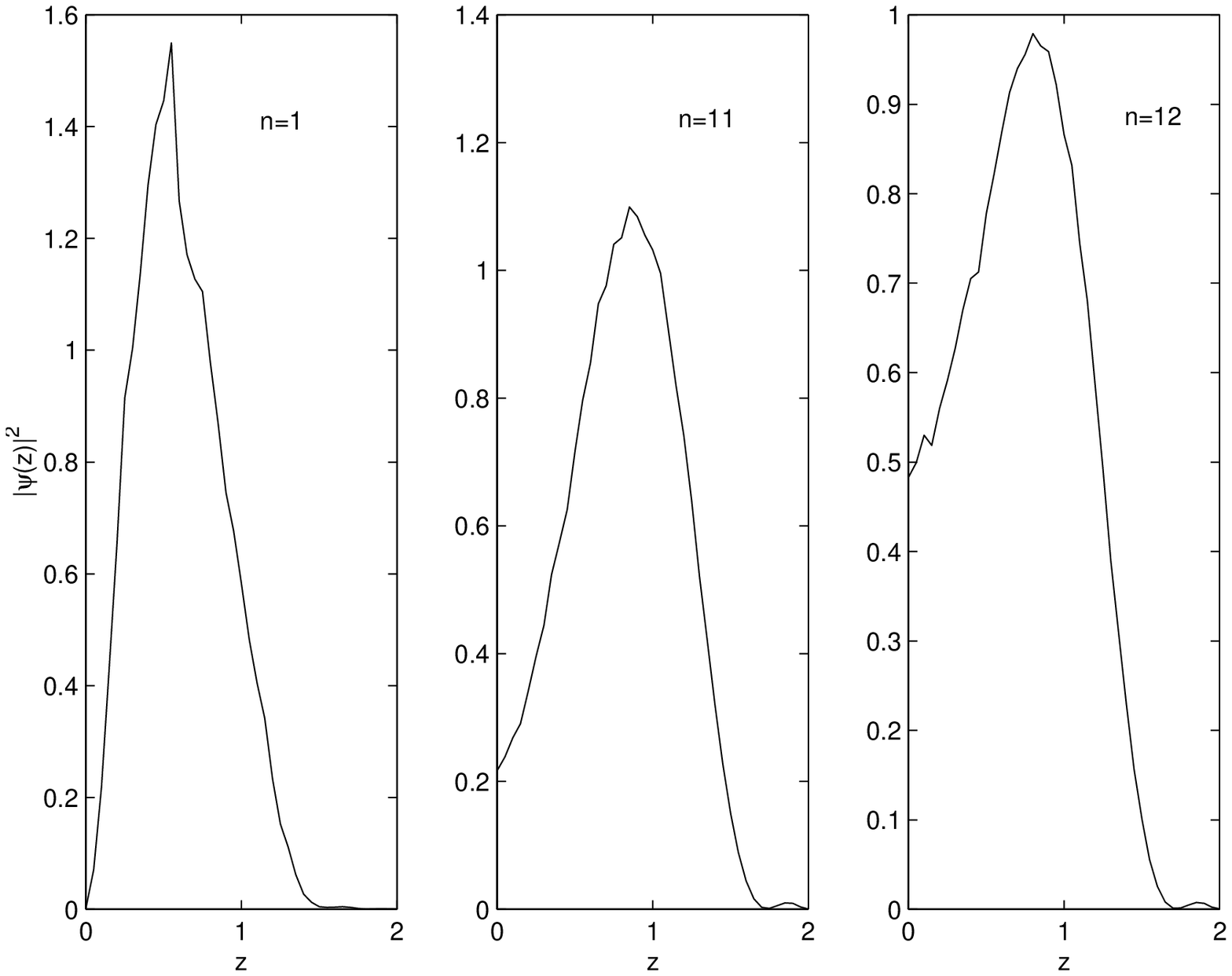,width=9truecm}
\end{center}
\caption{Projected wave functions for the ground state and levels 11 and 12}
\end{figure}

In all calculations, here and in the static case, care should be taken that
the basis that is chosen is not an orthogonal basis. Denote by $\chi
_{\alpha }$ the electron basis states defined in Eqs.(\ref{2.1}) - (\ref{2.2}%
). Let the eigenvector of the Hamiltonian corresponding to the eigenvalue $%
\lambda _{k}$ be 
\[
\phi ^{(k)}(\overrightarrow{x},z)=\sum_{\alpha n}c_{\alpha n}^{(k)}\chi
_{\alpha }(\overrightarrow{x},z)P_{n}(z)
\]
The coefficient $c_{\alpha n}^{(k)}$ is obtained from 
\[
c_{\alpha n}^{(k)}=\sum_{\beta m}V_{\alpha n,\beta m}\frac{1}{\sqrt{\gamma
_{\beta }}}A_{\beta m,k}
\]
$V$ being the matrix that diagonalizes the original basis and $A$ the matrix
that diagonalizes the Hamiltonian in the new orthonormalized basis. The
factor $\gamma _{\beta }$ is 
\[
\gamma _{\beta }=\sum_{\alpha \varepsilon }V_{\beta n,\alpha n}^{T}\left(
\chi _{\alpha },\chi _{\varepsilon }\right) V_{\varepsilon n,\beta n}
\]
Now the projected squared wave function on the $z-$axis is 
\[
\left| \psi ^{(k)}(z)\right| ^{2}=\sum_{\alpha n,\beta m}c_{\alpha
n}^{(k)*}c_{\beta m}^{(k)}P_{n}(z)P_{m}(z)\int_{\Sigma }\chi _{\alpha }(%
\overrightarrow{x},z)\chi _{\beta }(\overrightarrow{x},z)d^{3}x
\]
$\Sigma $ being the octahedron volume.

On sees from the lower plot in Fig.4 that, as expected, the ground state and
the first excited states correspond to a vanishing collision probability for
the positive particles. However after the 10th excited state many levels
appear that have a large value of the projected wave function at $z=0$.
These are {\it quantum collision states} which, when excited, lead the
positive particles to configurations of close proximity. In a energy
spectrum that, in the rescaled units of Eq.(\ref{2.5}) varies from 27.7 to
30000, the first states where $\left| \psi (0)\right| ^{2}$ is appreciably
different from zero lie only 26 units above the ground state. The meaning of
this energy difference in physical units will be discussed in the next
section. Fig.5 shows the projected wave functions for the ground state and
for levels 11 and 12.

\section{Conclusions and experimental implications}

Quantum collision states between like charged particles, in structures of
the type studied in the previous section, cannot be useful unless the
structure belongs to some lattice and a similar configuration is repeated
throughout, at least, a large part of the lattice. That is, the lattice must
act as a confining medium for the positive particle configurations.

To evaluate the potential usefulness of the quantum collision states for the
practical achievement of reactions between the positive particles, the first
step is to estimate, in physical units, the needed excitation energy when
the length parameter $a$ takes typical atomic values. For $a=2.05\stackrel{%
\circ }{A} $, the factor $\frac{4\pi \varepsilon _{0}a}{e^{2}}$ in Eq.(%
\ref{2.5}) implies that an excitation energy of $182$ $eV$ corresponds to
the $26$ units of the adimensional Hamiltonian $H^{^{\prime }}$ used in
Sect.2.2. That is, the excitation energies are in the high ultraviolet $-$
low X-ray range. These are energies at least one hundred times higher than
thermal excitations. Therefore one should no expect the quantum collision
states to be excited by thermal fluctuations. To make use of these states
electromagnetic radiation fields, in the high ultraviolet $-$ low X-ray
range, should be used.

After a very controversial start, the so called {\it cold fusion}
experiments were continued in several places in a much sober mood. Although
some of the initial claims could not be confirmed, there are undeniably a
few facts that defy chemical or thermodynamical explanations \cite{Nagel}.
Among them are the abnormal isotope ratios \cite{Arata, George} and a small
but well measured excess power effect \cite{Miles}.

Given the scale of energies needed to excite the quantum collision states,
discussed in this paper, it is not probable that they are at the origin of
the cold fusion events. It is much more probable that the observed events
arise from a mixture of exceptional causes. On the one hand, when deuterium
is absorbed by hydrogen storage materials, there is an expansion of the
crystal lattice and cracks are expected to occur. Strong electric fields may
occur in the cracks accelerating the deuterons to nuclear fusion energies.
This is consistent with the observation of protuberances and craters in cold
fusion samples, these being often the sites where unexpected elements appear
in high local concentrations.

On the other hand in hot spots caused by the (aggressive) electrolysis
process, instead of an ordered structure, one might have a hot soup of
electrons and deuterons and then, in this ergodic situation, it is known
that three body collisions DeD have a small but non-negligible probability 
\cite{Vilela2}. The occurrence of both situations, of course, will very much
depend on the material structure of the samples.

Being probably due to a mixture of exceptional events, it is therefore
natural for the cold fusion events to be hardly reproducible in any
controllable way. It is the opposite situation that would be surprising. By
contrast, the nuclear reaction control method that is being proposed here,
is supposed to operate only when a large number of lattice cells is occupied
by the right number of reactant nuclei and this in a regular manner to have
well defined excitation levels. It is therefore a method that, if feasible,
is fully controllable. A precondition for this scheme would be an accurate
experimental determination of the excitation levels, to design the
appropriate excitatory electromagnetic radiation field.

One might draw here a parallel with the current schemes for fusion by plasma
confinement. There, charged particles are hopefully confined by magnetic
fields, but confinement is not sufficient to achieve fusion. One needs to
heat the plasma by radiofrequencies. A metallic lattice is a much softer
confining device, but again one should not expect that just by confining the
particles, many energetically useful fusion events would take place. That
would be an unexpected miracle. To confine the deuterons, or other
reactants, on a lattice seems a sound approach, but then some other
mechanism, like the one discussed in this paper, must be found to induce the
desired reaction. It is this approach that here (and elsewhere \cite{Vilela3}%
) is called {\it hybrid fusion}.

Incidentally, both in the quantum collision states method and in the ergodic
situation (but not necessarily in crack acceleration) the dominant process
for the deuteron fusion would be three body DeD events, which would lead to
a preferred channel \cite{Vilela2}
\[
\begin{array}{lll}
D+e+D & \longrightarrow  & ^{4}He^{*}(20.1)+e \\ 
&  & \downarrow  \\ 
&  & \rightarrow T+p
\end{array}
\]

\end{document}